\documentclass[conference]{IEEEtran}
\IEEEoverridecommandlockouts
\usepackage{cite}
\usepackage{amsmath,amssymb}
\usepackage{graphicx}
\usepackage{booktabs}
\usepackage[T1]{fontenc}
\usepackage[utf8]{inputenc}
\usepackage{url}
\usepackage{placeins}
\usepackage[hidelinks]{hyperref}
\def\BibTeX{{\rm B\kern-.05em{\sc i\kern-.025em b}\kern-.08em T\kern-.1667em\lower.7ex\hbox{E}\kern-.125emX}}

\begin{document}

\title{Matching Workloads to Blackwell:\\
Power-Resolved Benchmarking of NVIDIA B200 and B300 across HPC and AI Workloads}

%
\author{\IEEEauthorblockN{Mohsen Zand\IEEEauthorrefmark{1},
Trung Dac Nguyen\IEEEauthorrefmark{1}, H. Birali Runesha}
\IEEEauthorblockA{Research Computing Center, University of Chicago, Chicago, IL 60637, USA\\
\{zand, ndtrung, runesha\}@uchicago.edu}
\thanks{\IEEEauthorrefmark{1}These authors contributed equally.}}

\maketitle

\begin{abstract}
We evaluate NVIDIA B200 and B300 (Blackwell Ultra) GPUs on two independent eight-GPU clusters, spanning computer-vision inference and training, LLM inference with Llama-3-8B and 70B~\cite{llama3}, and six production HPC codes. 
The B300 is not a uniform successor to the B200: it trails by up to 5.6$\times$ on convolutional inference, 8$\times$ on 3D vision transformer training, 24--27\% on 70B LLM throughput, and 16$\times$ on double-precision Quantum ESPRESSO. 
We resolve the convolutional deficit using per-GPU board power rather than throughput alone: on the affected workloads the B300 draws only 36--53\% of its 1100\,W envelope, whereas on ViT-Large, which it wins, both parts saturate at 95--97\% of TDP. 
Because our harness keeps inputs resident in device memory, a host-side data pipeline is excluded by construction, localizing the cause to per-kernel cost. 
Two B300 nodes of identical silicon differ by 2.1$\times$ in time-to-first-token solely from a development vLLM build. 
We give per-workload procurement guidance.
\end{abstract}


\section{Introduction}
NVIDIA's Blackwell generation has been positioned as a uniform advance over Hopper, and the B300 (Blackwell Ultra) as an advance over the B200: 288\,GB of HBM3e against 192\,GB, and roughly 8\,TB/s of memory bandwidth. 
Research computing centers must convert such specifications into procurement decisions, yet application-level speedup is governed by memory access patterns, launch overhead, collective communication, and above all by the maturity of the vendor software stack at the moment of purchase.

This paper reports what we measured when we ran our center's representative workloads on both parts. 
Our contributions are:

\begin{itemize}
\item A cross-workload evaluation of B200 and B300 on \emph{two} independent clusters, covering computer vision (CV) inference and DDP training, LLM inference at two model scales and three precisions, a 3D vision transformer, and six HPC codes (Section~\ref{sec:ai}, Section~\ref{sec:hpc}).
\item A power-resolved diagnosis of the B300 convolutional deficit. Throughput alone identifies that the B300 is slower; per-GPU board power identifies \emph{why}, and excludes the explanation we first assumed (Section~\ref{sec:power}).
\item Quantification of how much of an apparent hardware gap is in fact build configuration, using two nodes of identical silicon running different vLLM builds (Section~\ref{sec:llm}).
\end{itemize}

\section{Related Work}
Vendor-published figures report peak throughput or MLPerf-style results on a curated workload set~\cite{mlperf}, which bound what an application can achieve but not what it does achieve. 
The established way to close that gap is microbenchmarking: Luo et al.~\cite{luohopper1,luohopper2} dissect the Hopper architecture at instruction and memory-hierarchy level, and Jarmusch et al.~\cite{blackwellmicro} extend the same treatment to Blackwell, though on consumer silicon (a GeForce RTX 5080 against an H100 PCIe) rather than the datacenter parts. 
That literature characterizes the device; it does not tell a computing
center what a production code will do on it, and it does not cover B200 or B300.

Our study is complementary and differs in three respects. First, it spans AI and traditional HPC codes under a single methodology, where microbenchmark studies isolate architectural primitives. 
Second, it uses board power as a diagnostic instrument rather than only as an efficiency metric, which is what allows a device that is working hard and losing to be separated from one that is idle between kernels. 
Third, it replicates every AI measurement on two independently administered clusters, which is what allows software-stack effects to be separated from silicon. 
To our knowledge, no published evaluation reports independent B200-versus-B300 data across both AI and HPC workloads.

\section{Experimental Setup}
\label{sec:setup}
We use two eight-GPU-per-node Blackwell clusters, one from Lenovo and one from Schmidt Sciences, with A40, A100, H100 and H200 baselines drawn from Midway, the prior-generation ecosystem at our center. 
Both Blackwell clusters run CUDA~13.2; driver versions and host allocations differ and are tabulated in Appendix~\ref{app:ad}. 
Driver-reported HBM3e is 179\,GiB per B200 and 269\,GiB per B300, which are the vendor's
192\,GB and 288\,GB expressed in binary units, at 1000\,W and 1100\,W TDP respectively.

CV benchmarks use PyTorch~\cite{pytorch} with \texttt{torchrun} for DDP. 
Input tensors of shape $3\times224\times224$ are allocated once directly in device memory and reused across iterations, which isolates device and kernel performance from storage, decode and host-to-device transfer at the cost of excluding input-pipeline effects. 
Inference reports 100 timed iterations after 30 warmup iterations; training reports 50 after
15. The inference sweep covers all four models; because DDP training is substantially more
expensive per configuration, the training sweep covers ResNet-50 and ViT-Large only, chosen as
the launch-heavy and large-GEMM endpoints of the four. 
Board power is sampled at 20\,Hz through NVML on rank~0 across each timed region and reported as the mean.
Replication differs by workload class, and we state it explicitly. 
The CV and LLM figures are \emph{single} timed runs of the iteration counts above; spot checks put run-to-run variation at 2--4\%, and we draw no conclusion from differences below that band. 
The 3D ViT results (Table~\ref{tab:pangu}) and the HPC results (Tables~\ref{tab:hpcapps} and~\ref{tab:namd}) are \emph{medians of five} repeated runs.
Table~\ref{tab:power}, which carries the central diagnostic argument, therefore derives from single-run CV measurements, with power reported as a 20\,Hz NVML mean over each timed region. 
LLM benchmarks use vLLM~\cite{vllm} with tensor parallelism, 150 prompts spanning short, medium and long lengths, and greedy decoding capped at 256 tokens. 
The harness is a single driver that sweeps models, precisions, GPU counts and batch sizes, writing one CSV row per configuration; it is released with the paper (Section~\ref{sec:repro}).

Because our leading explanation concerns library maturity, we record the stack exactly.
Both clusters ran CUDA 13.2 with PyTorch 2.9.0, torchvision 0.24.0, cuDNN 9.15.1.9, cuBLAS 13.1.0.3, NCCL 2.28.9, Triton 3.5.0 and Python 3.11.14. 
Driver versions were 595.58.03 on Lenovo B200 and 595.71.05 elsewhere. 
LLM runs used vLLM 0.11.2 on Schmidt and a development build (\texttt{v0.1.dev0}) on Lenovo B300. 
\textbf{\texttt{torch.backends.cudnn.benchmark} was left at its default of \texttt{False}}, so convolution algorithms were chosen by cuDNN heuristics rather than by exhaustive autotuning. This is a detail that matters here, because heuristic selection is precisely what is most likely to be mis-tuned on a new architecture.
Board power is sampled on rank~0 only; per-GPU efficiency figures therefore assume the other seven GPUs draw comparably, which is reasonable under the symmetric DDP and tensor-parallel decompositions used here but which we did not verify directly.

\section{AI/ML Workloads}
\label{sec:ai}

\subsection{Computer vision}
Across four torchvision models --- ResNet-50~\cite{resnet}, EfficientNet-B4~\cite{effnet}, ViT-Large~\cite{vit} and ConvNeXt-Base~\cite{convnext} --- at 1, 4 and 8 GPUs with batch sizes from 64 to 8192, the B200 leads decisively on convolutional networks and loses on the transformer. 
At 8 GPUs in BF16, ResNet-50 reaches 110.7\,K\,FPS on B200 against 35.6\,K on B300 (3.1$\times$), and ConvNeXt-Base 52.9\,K against 9.4\,K (5.6$\times$). 
EfficientNet-B4 is nearly level (45.8\,K vs.\ 39.0\,K). 
ViT-Large reverses the ordering entirely: 47.7\,K\,FPS on B300 against 39.5\,K on B200, a 21\% advantage. 
DDP training reproduces the pattern: ResNet-50 attains 20.3\,K\,img/s on B200 against 6.8\,K on B300, while ViT-Large favors the B300 at 10.6\,K against 9.1\,K (+16\%). 
Per-model curves are given in Appendix~\ref{app:cv}.

BF16 and FP16 agree to within 6\% on ResNet-50, ViT-Large and ConvNeXt-Base, consistent with the two formats sharing tensor-core throughput on Blackwell. EfficientNet-B4 does not follow: FP16 leads BF16 by 21--27\% on the B200 while BF16 leads by 13.7\% on the B300, a reversal reproduced on both clusters at every GPU count and batch size, which points to kernel selection rather than to the formats themselves. We report BF16 throughout unless stated otherwise; the full precision sweep is in the released CSVs.

Two observations constrain any explanation. 
First, the deficit persists across the entire batch sweep (Appendix~\ref{app:cv}, Fig.~\ref{fig:cv_batch}), which is inconsistent with a per-batch overhead and consistent with a per-kernel cost. 
Second, agreement between clusters is markedly asymmetric: the two B300 nodes reproduce one another to within 1.1\% on every CV model --- and to within 0.13\% on the three convolutional models --- whereas the two B200 nodes differ by as much as 14\% (ViT-Large: 39.5 vs.\ 44.9\,K\,FPS). 
A part pinned to identical throughput regardless of host configuration, while its predecessor varies with it, indicates a limit invariant across hosts rather than one set by host resources.

\subsection{Power-resolved diagnosis}
\label{sec:power}
\begin{figure}[t]
  \centering
  \includegraphics[width=\linewidth]{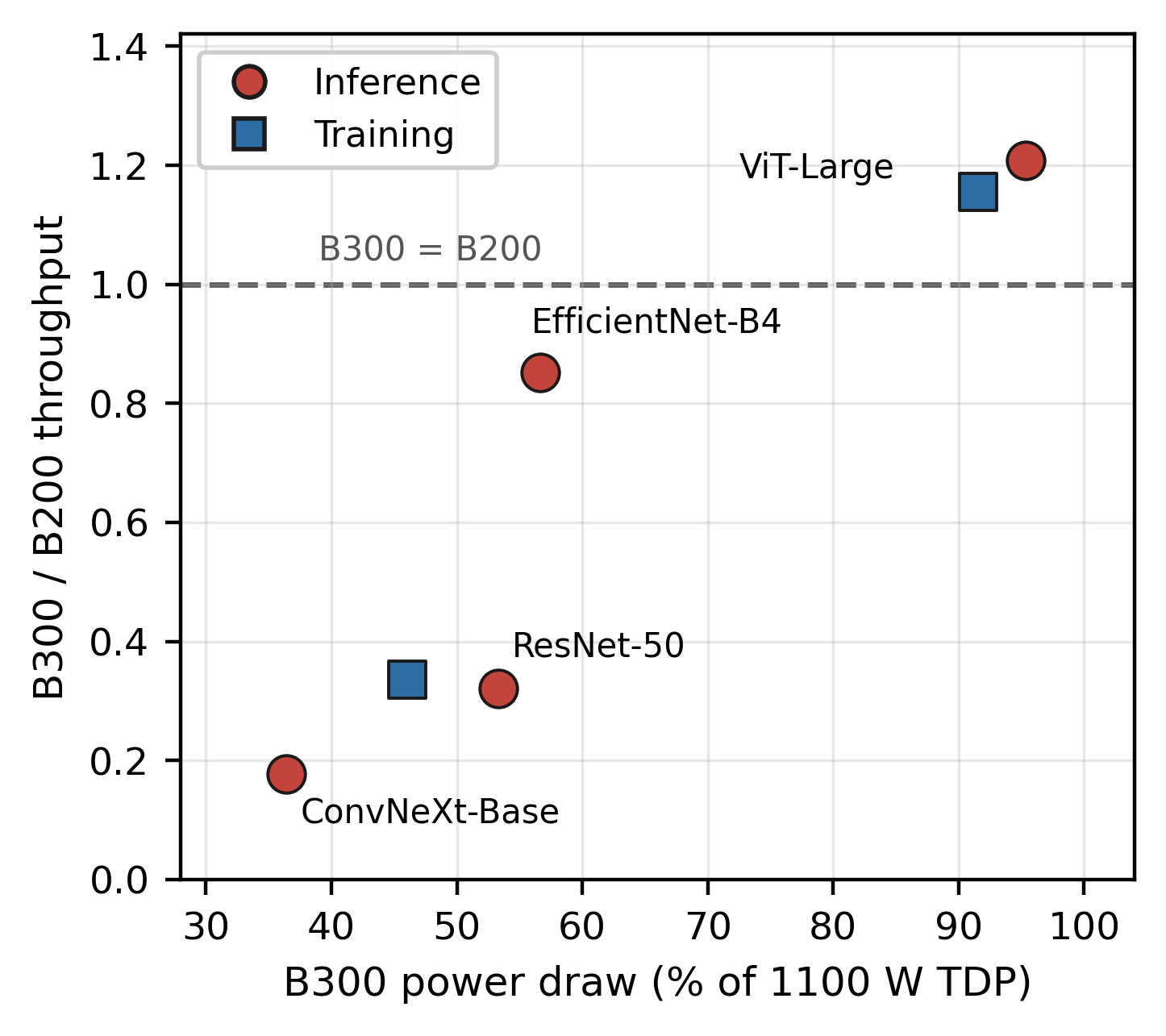}
  \caption{The diagnosis in one view. Each point is one model at its peak-throughput configuration (8 GPUs, BF16); training covers two of the four models (Section~\ref{sec:setup}). 
  The B300's throughput relative to the B200 rises monotonically with how much of its 1100\,W envelope it draws. 
  The two workloads it wins are the two that saturate the device; the workload it loses worst draws barely a third of available power.}
  \label{fig:diagnostic}
\end{figure}

Throughput alone cannot separate a device that is working hard and losing from one that is idle. 
Board power can. Table~\ref{tab:power} reports draw at the peak-throughput configuration as a fraction of each part's TDP, and Fig.~\ref{fig:diagnostic} plots the resulting relationship directly.

\begin{table}[t]
\caption{Per-GPU board power at peak-throughput configuration (8 GPUs, BF16, Lenovo), as a fraction of TDP (1000\,W B200, 1100\,W B300), against the B300/B200 throughput ratio. 
The deficit tracks the idle fraction monotonically.}
\label{tab:power}
\centering
\begin{tabular}{lccc}
\toprule
Model & B200 \%\,TDP & B300 \%\,TDP & B300/B200 \\
\midrule
\multicolumn{4}{l}{\textit{Inference}}\\
\quad ViT-Large       & 97 & 95 & \textbf{1.21} \\
\quad EfficientNet-B4 & 65 & 57 & 0.85 \\
\quad ResNet-50       & 80 & 53 & 0.32 \\
\quad ConvNeXt-Base   & 76 & \textbf{36} & 0.18 \\
\midrule
\multicolumn{4}{l}{\textit{Training (DDP)}}\\
\quad ViT-Large       & 93 & 92 & \textbf{1.16} \\
\quad ResNet-50       & 58 & 46 & 0.34 \\
\bottomrule
\end{tabular}
\end{table}

The ordering is monotonic. ViT-Large, the one workload on which the B300 wins, is also the one on which both parts saturate their power envelope; it wins on energy efficiency as well (5.68 vs.\ 5.11\,img/J in inference). 
At the other extreme, ConvNeXt-Base drives the B300 to just 36\% of its envelope while delivering 5.6$\times$ less throughput.

Energy efficiency follows the same split (Appendix~\ref{app:ad}). 
Per GPU, the B200 delivers 17.3\,img/J on ResNet-50 against 7.6 on the B300, and 8.8 against 2.9 on ConvNeXt-Base; a 2.3$\times$ and 3.0$\times$ advantage respectively. 
ViT-Large again inverts: 5.68\,img/J on B300 against 5.11 on B200. 
For a site paying for both the part and its power, the convolutional case is therefore worse than the throughput numbers alone suggest, because the B300 loses on throughput while its higher envelope raises the denominator.

A device that is compute- or bandwidth-bound saturates its power budget; a device drawing a third of it is idle between kernels. 
Because our harness holds inputs resident in device memory (Section~\ref{sec:setup}), a host-side data-transfer bottleneck is excluded by construction; an explanation we initially assumed and the measurement refuted.

Two candidates remain: CPU-side kernel-launch overhead, which dominates for launch-heavy networks of many small convolutions, and the maturity of sm\_103 convolution kernels in cuDNN~\cite{cudnn} 9.15. 
ResNet-50 and ConvNeXt-Base are launch-heavy; ViT-Large is a small number of large GEMMs. 
One piece of evidence favors the second: convolution algorithms were selected by cuDNN heuristics rather than autotuned (Section~\ref{sec:setup}), and heuristic tables are among the last things to be tuned for a new compute capability. 
That is suggestive, not conclusive.

We are explicit that our data do not separate the two, and we name the discriminating experiment rather than defer it vaguely. 
Capturing the forward pass in a CUDA graph, or compiling with \texttt{torch.compile} in \texttt{reduce-overhead} mode, removes essentially all
per-launch cost: a deficit that survives capture excludes launch overhead, and one that disappears confirms it. 
An \texttt{nsys} timeline reporting kernel durations against inter-kernel gaps would settle it equally directly. 
Neither requires new hardware, only scheduled access to both parts, which we did not have before this version. 
We flag it as the single measurement that would close the diagnosis.

\subsection{LLM inference}
\label{sec:llm}
For Llama-3-8B at 8 GPUs, the B200 leads by 16\% on Lenovo (24.2 vs.\ 20.9\,K\,tok/s); for Llama-3-70B, it leads by 24--27\% (11.0--11.2\,K vs.\ 8.7--9.0\,K). 
Tensor parallelism scales poorly at 8B, yielding only 12--30\% additional throughput for 8$\times$ the hardware, since TP targets latency rather than aggregate throughput; throughput-oriented deployments at this scale should prefer data-parallel replicas. 
At 70B, scaling reaches 2.5$\times$.

The precision results expose a software effect large enough to be mistaken for hardware. 
On B200 at 70B, FP8 outperforms FP16 by 34--36\%, yet BF16 still leads at 11.0\,K\,tok/s. 
Since BF16 and FP16 share tensor-core throughput on Blackwell, that gap is not architectural: it is absent at 1 GPU (4.1 vs.\ 4.1\,K\,tok/s) and widens with TP width to 38\% at 8 GPUs, indicating that the FP16 path selects a less efficient collective or attention backend under tensor parallelism.

At 8B, the precision ordering is inverted between parts. 
The B200 attains its highest throughput in BF16 (23--24\,K\,tok/s), whereas on the B300 FP16 marginally leads BF16 (Schmidt 23.8 vs.\ 22.9\,K; Lenovo 21.6 vs.\ 20.9\,K); a margin inside the 2--4\% run-to-run band and not one we interpret. 
FP8 is the substantive case: it lifts Schmidt B300 to 22.7\,K\,tok/s against 19.0\,K on Schmidt B200, a 19\% advantage that is the clearest evidence of Blackwell Ultra's low-precision orientation.

The clearest demonstration is cross-cluster. The two B300 nodes are identical silicon, but Lenovo runs a development vLLM build (\texttt{v0.1.dev0}) and Schmidt a production release.
Schmidt B300 gains 19\% from FP8 at 8B; Lenovo gains 7\%. 
Lenovo B300 time-to-first-token is 2.1$\times$ higher at 8B and 1.6$\times$ higher at 70B than Schmidt B300, from less optimized CUDA-graph compilation for sm\_103. 
Any procurement conclusion drawn from the Lenovo node alone would have misattributed a build difference to the architecture. 
In absolute terms, the sub-10\,ms (8B) and sub-15\,ms (70B at $\geq$4 GPUs) interactive responsiveness targets our users require are met by the B200 and by Schmidt B300, but not by Lenovo B300 pending a production sm\_103 release. 
Full precision and TTFT results are in Appendix~\ref{app:llm}.

\subsection{3D vision transformer}
Training Pangu-Weather~\cite{pangus2s} at a patch size of $2\times4\times4$, using a training code base released by our group\footnote{\url{https://github.com/ndtrung81/3dViT-training}}, the B200 sustains 8.3 samples/s versus 1.0 on the B300 at the optimal global batch size of 12 over 8 GPUs.
Performance peaks at 4 GPUs and regresses at 8, and FP8 on the B300 is slower than BF16, which we attribute to the same immature-stack effect. See Appendix~\ref{app:pangu}.

\section{HPC Workloads}
\label{sec:hpc}
Six production codes were run with vendor-recommended configurations; the full result tables are in Appendix~\ref{app:hpc}.

\textbf{Double precision is where the B300 is weakest.} 
For LAMMPS~\cite{lammps} with KOKKOS~\cite{kokkos} on the Lennard-Jones benchmark, B300 double-precision performance is 2.6--2.7$\times$ \emph{lower} than H200 across all five system sizes; B300 single-precision merely matches H200 double-precision. 
For Quantum ESPRESSO~\cite{qe2009,qe2017}, which requires double precision throughout, we ran the $G_0W_0$ calculation on a 511-atom NV-centre diamond supercell; its dominant kernel, $h\_psi$, runs 16$\times$ slower on a B300 than on a B200. 
The same kernel on the B200 matches an H200 and beats an A100 by 2.6$\times$.

We hold this explanation to the same evidentiary standard as the convolutional one. 
NVIDIA specifies HGX B200 at 296\,TFLOP/s of FP64 against 10\,TFLOP/s for HGX B300~\cite{hgxspec} (dense figures: the sparsity footnotes on that specification cover only the FP4--TF32 rows), i.e.\ roughly 37 against 1.25\,TFLOP/s per GPU, a nominal ratio near 30$\times$. 
The measured Quantum ESPRESSO gap of 16$\times$ is therefore about half the specification ratio. 
That the measured deficit falls \emph{below} the nominal one is what the architectural account predicts: the calculation is not purely FP64-throughput-bound, so it recovers part of the difference elsewhere. 
Had the measured gap \emph{exceeded} the specification ratio we would have had to look beyond die area for an explanation; it does not. 
The LAMMPS double-precision comparison is against an H200 rather than a B200 and so is not directly commensurable with this ratio, but its much smaller 2.7$\times$ gap points the same way: the Lennard-Jones benchmark is bound substantially by something other than double-precision throughput.

\textbf{Single-precision codes behave as expected.} GROMACS~\cite{gromacs} on the STMV benchmark runs about 2$\times$ faster on 4$\times$B200 or 4$\times$B300 than on 4$\times$H200, and 4.4$\times$ faster than 4$\times$A100; for this one-million-atom system the optimal configuration places one MPI rank per GPU with four to eight threads per rank, and GROMACS uses single precision throughout on both host and device. 
NAMD~\cite{namd} gains 2.2--2.7$\times$ over A100 across 67K--333K atoms, and OpenMM~\cite{openmm} 1.8--2.4$\times$ across 25K--277K atoms; these systems are too small to saturate either part, so the speedup tracks the memory-bandwidth ratio. 
In NAMD 3.0 GPU-resident mode~\cite{namd3} on a 679K-atom system, the B200 leads the B300 by roughly 1.2$\times$ at 1, 2 and 4 GPUs, with the optimal number of processing elements per GPU falling as GPU count rises (16, 16 and 8 respectively) --- the scaling is clean, but the ordering between parts does not change.

\textbf{ML-driven scientific codes are mixed.} 
Gen-COMPAS~\cite{gencompas} diffusion training gains only 1.4$\times$ on B200 over A100, and on B300 \emph{regresses} to 9.3 iterations/s, 3.2$\times$ below the A100; the same signature as the convolutional deficit.

We state plainly what the B300 HPC guidance rests on. Table~\ref{tab:hpcapps} compares A100 against B200 only: we did not have B300 allocation for the NAMD, OpenMM and LAMMPS-MACE benchmarks. 
The B300 row in Table~\ref{tab:rec} is therefore supported by Quantum ESPRESSO, LAMMPS double precision, NAMD in GPU-resident mode (Table~\ref{tab:namd}) and Gen-COMPAS, and not by the single-GPU molecular-dynamics sweep.

\section{Discussion and Recommendations}
Two conclusions follow that a specification sheet would not have predicted.

First, the B300 should be purchased for what it is: a part optimized for large-GEMM, low-precision transformer work with a large weight footprint. 
It is the better choice for ViT-class inference and training, where it wins on both throughput and energy efficiency, and for workloads gated on per-device capacity. 
It is the worse choice for convolutional networks, for FP64 codes, and --- at present --- for 70B LLM serving. 
Sites running mixed workloads should not treat B300 as a drop-in successor.

For HPC practitioners specifically, three consequences follow. 
A site whose load is dominated by double-precision electronic-structure or implicit-solver work should not buy B300 at all: at roughly 1.25\,TFLOP/s of FP64 per GPU it is slower than hardware two generations older, and no software update will change that. 
A site running single-precision molecular dynamics should decide per code rather than per class: GROMACS is interchangeable between the two parts to within 1\%, so price, power and capacity decide; NAMD in GPU-resident mode is not, giving the B200 a 1.2$\times$ lead (Table~\ref{tab:namd}); and OpenMM we could not measure on the B300. 
A site running PyTorch-based scientific ML --- surrogate models, diffusion samplers, neural potentials --- should profile kernel granularity on a representative step before committing, because that is where we observe the largest and least expected regressions, including a case where the B300 falls behind an A100.

A synthesis across the two halves of the study sharpens this. 
We first read the deficit as CNN-specific, because that is where it is largest. It is not. 
The 3D vision transformer regresses by 8$\times$ on the B300 despite being a transformer, and Gen-COMPAS diffusion training --- PyTorch, BF16, unrelated to vision --- falls to 3.2$\times$ \emph{below} an
A100. 
What these share with ResNet-50 and ConvNeXt-Base is not their layer type but their execution profile: many small kernels launched per step, with no single operation large enough to saturate the device. 
ViT-Large inference, by contrast, issues a few large GEMMs and saturates. 
EfficientNet-B4 is the instructive intermediate case: it is a convolutional network, yet it reaches 0.85 of B200 throughput at 57\% of TDP, sitting between the two extremes on both axes rather than with the other CNNs. 
Its depthwise-separable design issues fewer and larger kernels than ResNet-50 or ConvNeXt-Base, and its result tracks that fact rather than its model family. 
The predictive variable is kernel granularity, not model family, and it is visible in the power trace before it is visible in any throughput number. 
Sites evaluating Blackwell Ultra for training workloads should profile kernel granularity rather than extrapolate from a model's architectural label.

Second, at this point in an architecture's life cycle, build maturity is a first-order performance variable, not a footnote. 
A single-cluster evaluation of the B300 would have reported a 2.1$\times$ TTFT penalty and a 7\% FP8 gain as properties of the silicon; the second cluster shows both to be properties of a vLLM build. 
We therefore recommend that procurement benchmarking of a new architecture either use two independently administered systems, or pin and report the exact build provenance of every component in the stack.

\begin{table}[t]
\caption{Per-workload procurement guidance. Ratios are B300 relative to B200 across the configurations measured, except where noted.}
\label{tab:rec}
\centering
\begin{tabular}{llc}
\toprule
Workload class & Prefer & B300/B200 \\
\midrule
ViT-class inference / training & \textbf{B300} & 1.16--1.21 \\
Large per-device weight footprint & \textbf{B300} & capacity \\
Launch-heavy CNNs (ResNet-50, ConvNeXt) & B200 & 0.18--0.34 \\
EfficientNet-B4 & B200 & 0.85 \\
LLM serving, 8B and 70B & B200 & 0.79--0.98 \\
3D ViT (Pangu-Weather) & B200 & 0.12 \\
FP64, Quantum ESPRESSO & B200 & 0.06 \\
FP64, LAMMPS$^{\ast}$ & B200 & 0.37--0.38 \\
Single-precision MD (GROMACS) & either & $\approx$1.0 \\
\bottomrule
\end{tabular}
\\[2pt]
\footnotesize $^{\ast}$LAMMPS double precision was measured against the H200, not the B200; this ratio is B300/H200.
\end{table}

\subsection{Threats to validity}
Three caveats bound these conclusions. First, replication is uneven, as set out in Section~\ref{sec:setup}: the HPC and 3D ViT figures are medians of five runs, while the CV and LLM figures are single runs with spot-checked variation of 2--4\%. 
The effects we build conclusions on span 3$\times$ to 16$\times$, far outside that band, so the ordering is not in question; but the precision-level comparisons, where margins are near 4\%, should be treated as indicative until repeated. 
We have deliberately declined to interpret any difference below that threshold.

Second, our CV harness deliberately excludes the input pipeline. 
A deployment feeding real images from storage may well encounter a genuine data-supply bottleneck that our measurement is blind to by construction. 
Our claim is narrower than it may appear: not that input pipelines never limit these workloads, but that one cannot explain what we observe here, because there is no input pipeline in the timed region to limit anything.

Third, cross-generation comparisons against the prior-generation baselines carry a compiler and MPI difference (NVHPC 24.1 versus 26.1) alongside the hardware difference, and the two Blackwell clusters differ in host thread allocation (112--172 threads on Lenovo against a 64-thread Slurm allocation on Schmidt). 
The NVHPC version difference was not a choice: release 24.1 was incompatible with the CUDA driver installed on the Blackwell servers. 
The A100 and H100 ratios should therefore be read as end-to-end
deployment ratios rather than isolated silicon comparisons. 
We consider this acceptable because it is also the comparison a procurement decision actually faces: a site buys a system and a software stack together, not a die.

\subsection{Reproducibility}\label{sec:repro}
The benchmark harness, the raw per-configuration CSV records underlying every figure and table, and the plotting scripts are available at \url{https://github.com/MohsenZand/GPUBenchmarking}; the 3D ViT training code is at \url{https://github.com/ndtrung81/3dViT-training}. 
Every number in this paper is traceable to a
single row of a released CSV, including the board-power samples, so that the diagnosis in Section~\ref{sec:power} can be checked independently of our interpretation of it.
Appendix~\ref{app:ad} describes the artifact.

\section{Conclusion}
Across CV, LLM, weather-model and HPC workloads on two independent clusters, the NVIDIA B300 is not a uniform improvement on the B200. 
Per-GPU power draw, rather than throughput alone, localizes its convolutional deficit to per-kernel cost and rules out data supply, and cross-cluster comparison shows a substantial share of the remaining gap to be software maturity rather than silicon. 
The deficit is governed by kernel granularity rather than model family, which makes it invisible to specification sheets and to any evaluation that reports throughput without power. 
Procurement decisions for Blackwell Ultra should be made per-workload, profiled at kernel granularity, and benchmarked against the exact software build that will be deployed --- on more than one system where that is possible. 
We are releasing the full record so that others can test whether the pattern we observe holds on their own workload mix.

\section*{Acknowledgment}
We thank NVIDIA, Schmidt Sciences and Lenovo for access to the hardware resources. 
TDN thanks Christopher Chipot, Kristina Herman, Jonathan Harris, Scott Kaiser, Harsha Mandayam Bharathi, Sanket Walujkar, Trayder Thomas for providing the input decks for the HPC applications. 
We thank the Hassanzadeh group at the University of Chicago for sharing the original version of the 3D Vision Transformer training code and datasets. TDN thanks Youzhi Yu for helping with optimizing and modifying the training code in preparation for Blackwell GPU benchmark runs.

\section*{Disclosure of Large Language Model Use}
The prose of this paper was revised with the assistance of a large language model. The model was additionally used to verify reported values against the raw result files from the authors' benchmark records. 
All experiments, measurements and analyses are the authors' own. 
The authors have verified all text, tables and figures against the source data and take full responsibility for the content.

\clearpage
\appendices

\section{Artifact Description / Artifact Evaluation}
\label{app:ad}

\subsection{Overview of contributions and artifacts}
The artifact comprises the benchmark harness used for all CV and LLM measurements, the raw per-configuration CSV records behind every figure and table in this paper, and the plotting scripts that regenerate the figures from those records. 
It is archived at \url{https://github.com/MohsenZand/GPUBenchmarking}. 
The 3D ViT training code is a separate artifact at \url{https://github.com/ndtrung81/3dViT-training}.

\subsection{Computational artifacts}
\texttt{src/cv.py} sweeps torchvision models over precision, GPU count and batch size in inference or DDP training mode, sampling board power through NVML on rank~0 at 20\,Hz across each timed region. 
\texttt{src/llm.py} drives vLLM over tensor-parallel widths and precisions with a fixed 150-prompt workload. 
Each writes one CSV row per configuration. The columns are
\texttt{Tag, Model, Dtype, GPUs, Batch, FPS, Lat\_ms, Power\_W, Img\_per\_Watt, VRAM\_GB} for CV and \texttt{Tag, Model, TP, Dtype, TPS, Avg\_Lat\_ms, TTFT\_ms, VRAM\_Util} for LLM.

Two caveats on the recorded columns, both of which affect how they should be read.
\texttt{Power\_W} is rank~0's board power, not the node total, so the per-GPU efficiency figures reported in this paper are computed as $(\mathrm{FPS}/N_{\mathrm{GPU}})/P$ rather than from the stored \texttt{Img\_per\_Watt} column, which divides aggregate throughput by single-GPU power and is therefore inflated at $N_{\mathrm{GPU}}>1$. 
\texttt{VRAM\_Util} reflects vLLM's \texttt{gpu\_memory\_utilization} preallocation setting ($\approx$0.87), not measured occupancy.

\subsection{Reproduction}
With the environment in \texttt{env.yaml} instantiated, \texttt{run\_bench.sh} reproduces the full sweep on an eight-GPU node. A single configuration can be reproduced directly, for example: \begin{quote}\ttfamily\small\raggedright torchrun -{}-nproc\_per\_node=8 src/cv.py\\ -{}-mode inference -{}-models convnext\_base -{}-dtype bf16 \end{quote} 
Expect approximately 24 GPU-hours for the complete CV and LLM sweep on one node. 
Figures are regenerated with \texttt{src/vis\_paper\_nospecs.py} and \texttt{src/fig\_diagnostic.py}, which read only the released CSVs, so every figure in this paper can be rebuilt without access to Blackwell hardware.

\subsection{System and library versions}
Table~\ref{tab:specs} gives the configuration of the systems described in Section~\ref{sec:setup}. 
The two B300 nodes are identical in silicon and differ only in driver revision, host allocation and vLLM build, which is what makes the software-maturity comparison in Section~\ref{sec:llm} possible. 
Note that \texttt{torch.backends.cudnn.benchmark} was left at its default of \texttt{False} throughout.

\begin{table}[htbp]
\caption{Hardware and software configuration. HBM3e is as reported by \texttt{nvidia-smi} (183{,}359\,MiB and 275{,}040\,MiB per GPU); the vendor's 192\,GB and 288\,GB are the same capacities in decimal units, and the eight-GPU totals match the 1.4\,TB and 2.1\,TB quoted for HGX B200 and HGX B300~\cite{hgxspec}. 
Host CPU threads are the full-node count on Lenovo and the Slurm allocation (\texttt{--cpus-per-task}) on Schmidt.}
\label{tab:specs}
\centering
\small
\setlength{\tabcolsep}{4pt}
\begin{tabular}{lllll}
\toprule
                  & \multicolumn{2}{c}{B200} & \multicolumn{2}{c}{B300} \\
\cmidrule(lr){2-3}\cmidrule(lr){4-5}
                  & Lenovo & Schmidt & Lenovo & Schmidt \\
\midrule
GPUs per node     & 8$\times$SXM & 8$\times$SXM & 8$\times$SXM6 & 8$\times$SXM6 \\
HBM3e per GPU     & 179\,GiB & 179\,GiB & 269\,GiB & 269\,GiB \\
TDP per GPU       & 1000\,W & 1000\,W & 1100\,W & 1100\,W \\
NVIDIA driver     & 595.58  & 595.71  & 595.71  & 595.71 \\
CUDA toolkit      & 13.2    & 13.2    & 13.2    & 13.2 \\
Host CPU threads  & 112     & 64      & 172     & 64 \\
Host RAM          & node    & 500\,GB & node    & 500\,GB \\
\bottomrule
\end{tabular}
\end{table}

\FloatBarrier
\subsection{Benchmark Methodology}
\label{app:method}
For inference, the reported batch size is per GPU and throughput is aggregated over ranks; for training, the reported batch size is global and divided across ranks. 
Board power is sampled on rank~0 only, so the absolute figure is one GPU's draw, not the node's; percentages of TDP in Table~\ref{tab:power} and energy efficiency figures are therefore normalized per GPU as $(\mathrm{FPS}/N_{\mathrm{GPU}})/P$ to remain comparable across GPU counts.

For B300 FP8, vLLM must be built from source with \texttt{TORCH\_CUDA\_ARCH\_LIST=`10.3'} to enable sm\_103 CUTLASS kernels. 
NV-FP4 is not benchmarked: the current vLLM release cannot load \texttt{modelopt}-quantized Llama-3 checkpoints due to a format mismatch. 
All vLLM runs preallocate \texttt{gpu\_memory\_utilization} at approximately 0.87, so reported VRAM utilization reflects that setting rather than the model's intrinsic footprint.

LAMMPS 11Feb2026 was built with KOKKOS in single, mixed- and double-precision with the CUDA backend under CUDA Toolkit 13.2. 
Compilers and MPI come from NVHPC 24.1 on the prior-generation cluster and 26.1 on the two Blackwell clusters, so cross-generation comparisons carry a toolchain difference alongside the hardware difference. 
GROMACS v2025.3 was run from the NGC container (\texttt{nvcr.io/nvidia/gromacs:v2025.3}) under Apptainer.

\FloatBarrier
\subsection{Computer Vision Results}
\label{app:cv}
\begin{figure*}[htbp]
  \centering
  \includegraphics[width=0.95\textwidth]{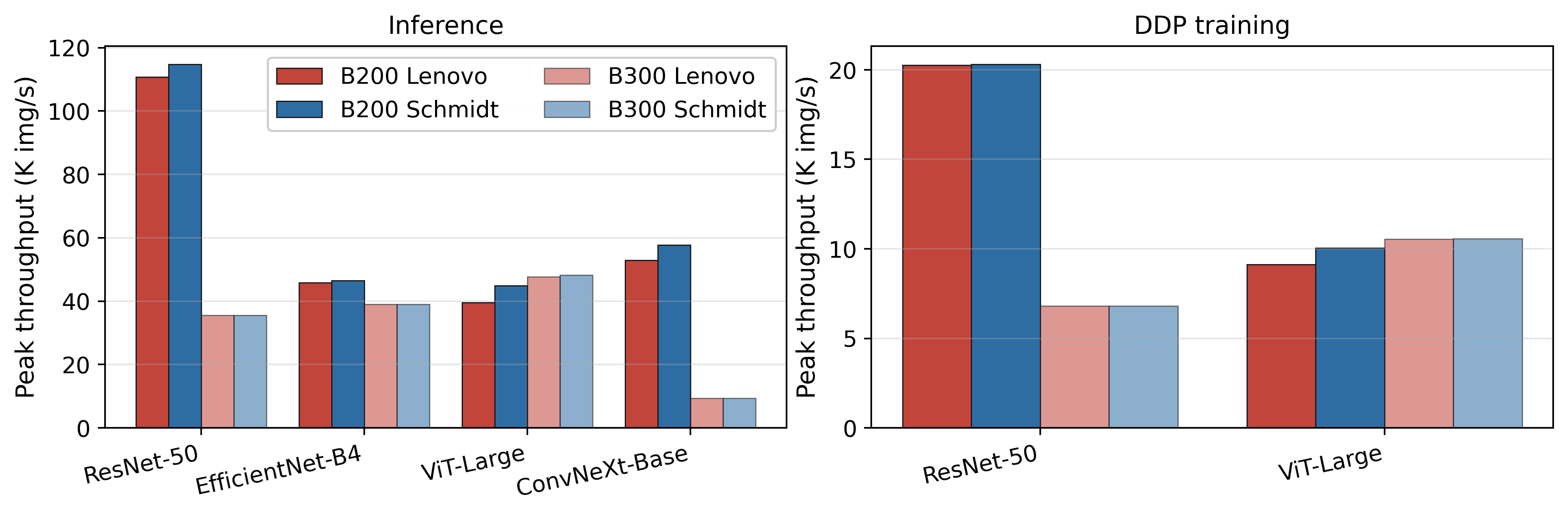}
  \caption{Peak CV throughput at 8 GPUs in BF16 on both clusters: inference over all four models (left) and DDP training over the two endpoint models (right, see Section~\ref{sec:setup}). 
  The two panels carry different vertical scales. The B300 trails on the convolutional models and leads on ViT-Large; the two B300 nodes agree to within 1.1\% on every model while the two B200 nodes differ by up to 14\%.}
  \label{fig:cv_scaling}
\end{figure*}

\begin{figure*}[htbp]
  \centering
  \includegraphics[width=0.95\textwidth]{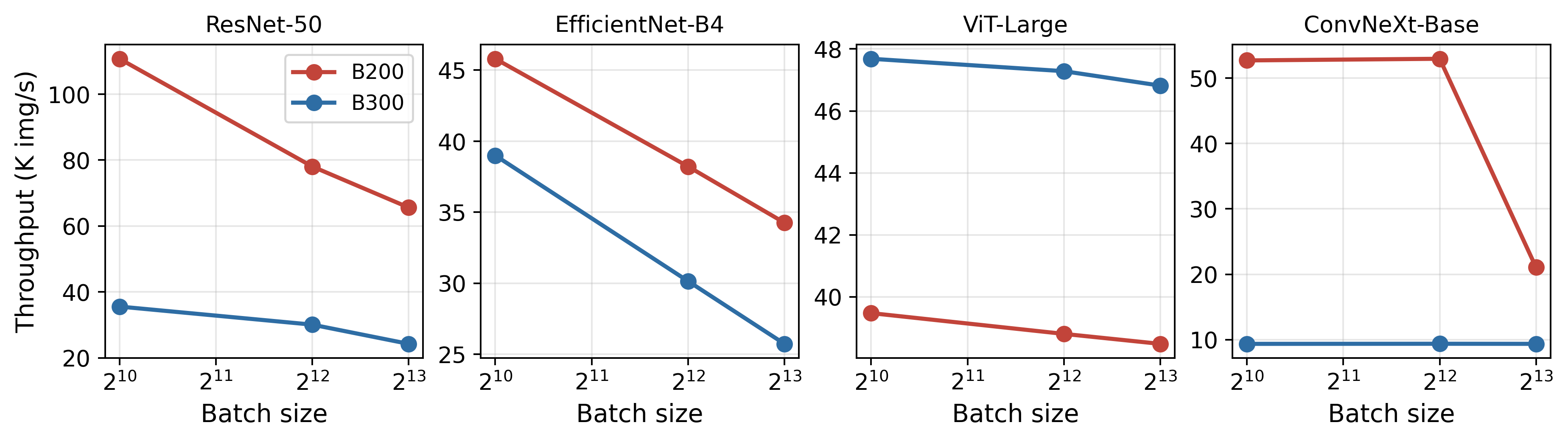}
  \caption{Throughput versus batch size at 8 GPUs (BF16, Lenovo). Each panel carries its own vertical scale, so read the gap within a panel rather than comparing heights across panels.
  The B300 convolutional deficit persists across the full sweep, which is inconsistent with a per-batch overhead and consistent with a per-kernel cost.}
  \label{fig:cv_batch}
\end{figure*}

\begin{figure*}[htbp]
  \centering
  \includegraphics[width=0.95\textwidth]{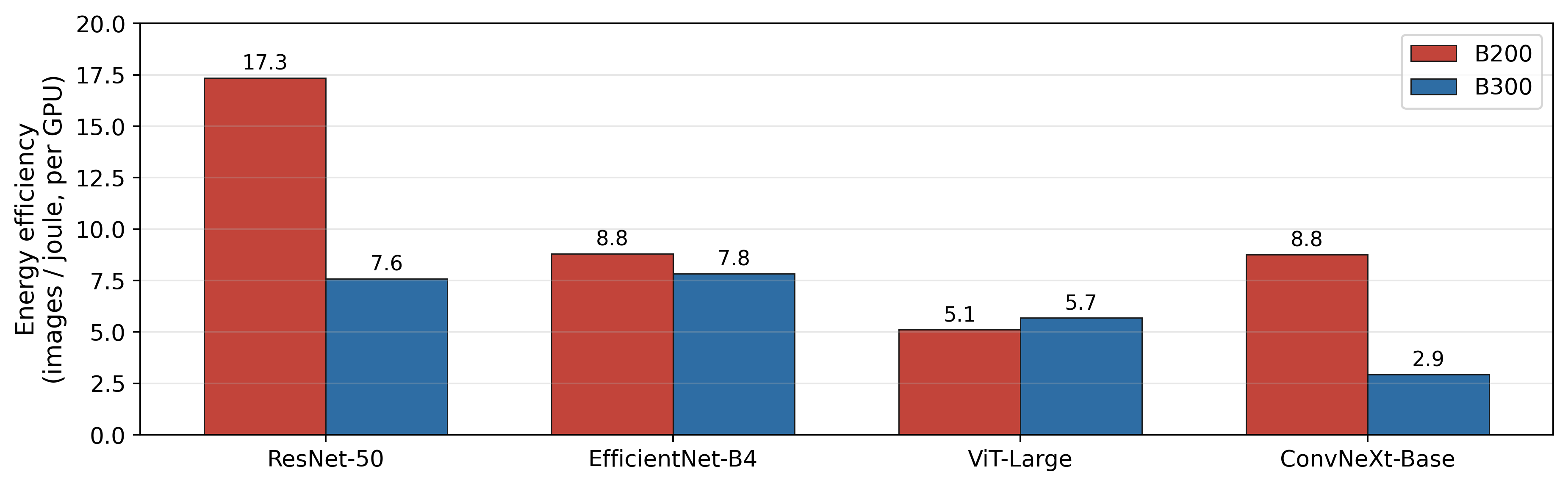}
  \caption{CV inference energy efficiency on the Lenovo cluster, from NVML board power sampled at 20\,Hz and normalized per GPU. ViT-Large is the only model on which the B300 is both faster and more efficient; on the three convolutional models it is worse on both axes despite the larger power envelope.}
  \label{fig:cv_energy}
\end{figure*}
These figures give the per-model detail behind Section~\ref{sec:ai}.
Fig.~\ref{fig:cv_scaling} shows peak inference and DDP training throughput;
Fig.~\ref{fig:cv_batch} the batch sweep, whose flatness rules out a per-batch overhead as the cause of the B300 deficit; and Fig.~\ref{fig:cv_energy} energy efficiency per GPU.

\begin{figure*}[htbp]
  \centering
  \includegraphics[width=0.95\textwidth]{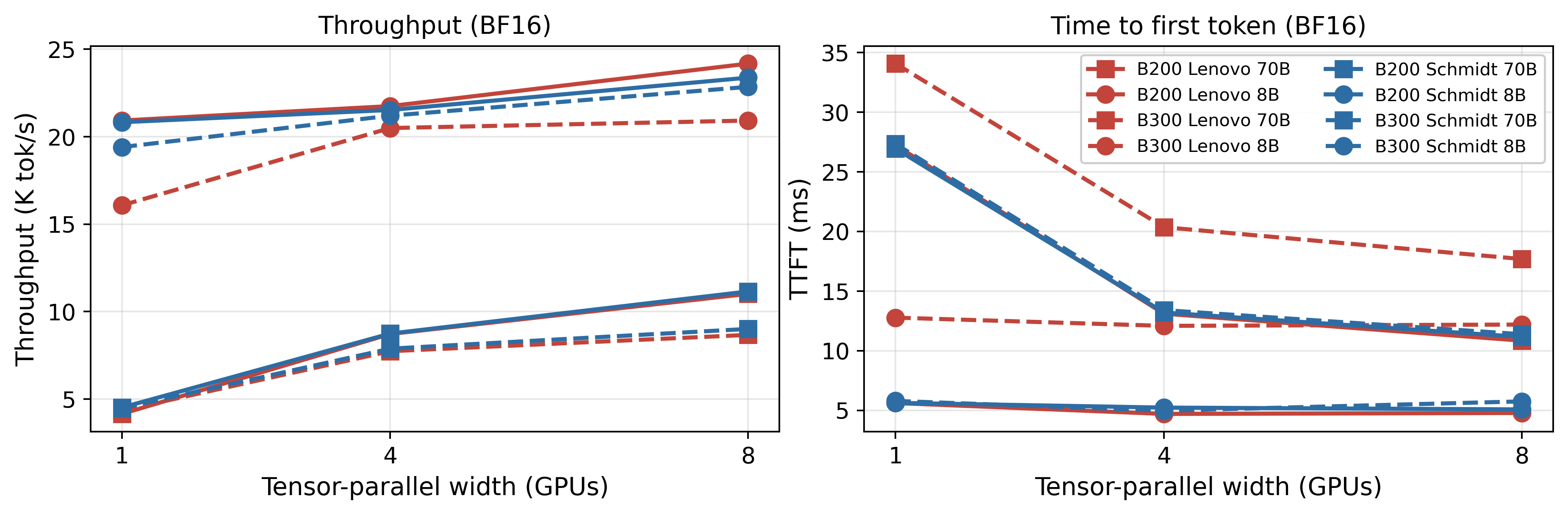}
  \caption{LLM throughput and time-to-first-token against tensor-parallel width (BF16). 
  Color distinguishes cluster (red Lenovo, blue Schmidt), line style distinguishes part (solid B200, dashed B300) and marker distinguishes model (squares 70B, circles 8B). 
  The separation between the two dashed 70B curves in the right panel --- B300 Lenovo at 34\,ms against B300 Schmidt at 27\,ms on one GPU --- is a vLLM build difference on identical silicon, not a hardware one.}
  \label{fig:llm_scaling}
\end{figure*}

\FloatBarrier
\subsection{LLM Results}
\label{app:llm}
These figures give the full LLM detail behind Section~\ref{sec:llm}, across tensor-parallel width and time-to-first-token (Fig.~\ref{fig:llm_scaling}). 
The separation between the two B300 curves is a vLLM build difference, not a hardware one.

\FloatBarrier
\subsection{3D Vision Transformer}
\label{app:pangu}
Table~\ref{tab:pangu} reports the factorial sweep over precision, global batch size and GPU count for Pangu-Weather training, summarized in Section~\ref{sec:ai}. 
Performance peaks at a global batch size of 12 and at four GPUs, regressing at eight. 
For this particular workload, we performed 5 repeated runs and reported the median value of the training performance.

\begin{table}[htbp]
\caption{Pangu-Weather 3D ViT training, samples/s.}
\label{tab:pangu}
\centering
\small
\begin{tabular}{llrrl}
\toprule
Parameter & Value & B200 & B300 & Constraints \\
\midrule
Precision & BF16 & 5.9 & 0.7 & batch 8, 8 GPUs \\
          & FP16 & 5.5 & 0.9 & \\
          & TF32 & 3.9 & 0.8 & \\
          & FP8  & n/s & 0.7 & \\
\midrule
Batch size & 8  & 5.9 & 0.7 & BF16, 8 GPUs \\
           & 12 & 8.3 & 1.0 & \\
           & 16 & 4.2 & 0.7 & \\
\midrule
GPUs & 2 & OOM & 1.5 & BF16, batch 16 \\
     & 4 & 4.5 & 1.7 & \\
     & 8 & 4.0 & 0.7 & \\
\bottomrule
\end{tabular}
\end{table}

\FloatBarrier
\subsection{HPC Application Results}
\label{app:hpc}
\begin{figure*}[htbp]
  \centering
  \includegraphics[width=0.78\textwidth]{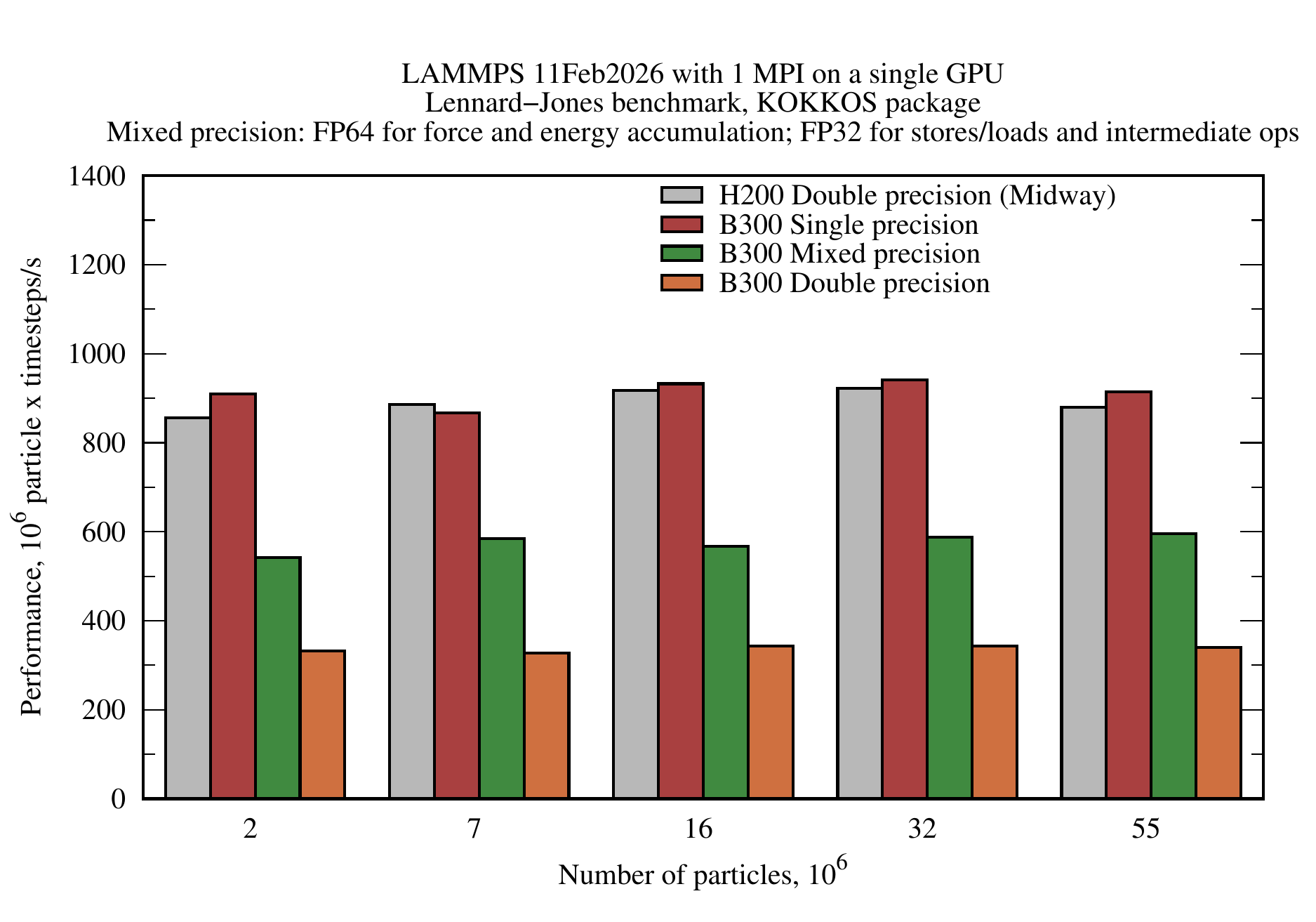}
  \caption{LAMMPS 11Feb2026 (KOKKOS) Lennard-Jones throughput versus problem size, single GPU. 
  Three B300 precisions are shown against H200 double precision: B300 double precision trails the H200 by 2.6--2.7$\times$ at every size, and B300 \emph{single} precision only draws level with H200 \emph{double} precision. 
  Mixed precision keeps FP64 for force and energy accumulation and FP32 elsewhere.}
  \label{fig:lammps}
\end{figure*}

\begin{figure*}[htbp]
  \centering
  \includegraphics[width=0.62\textwidth]{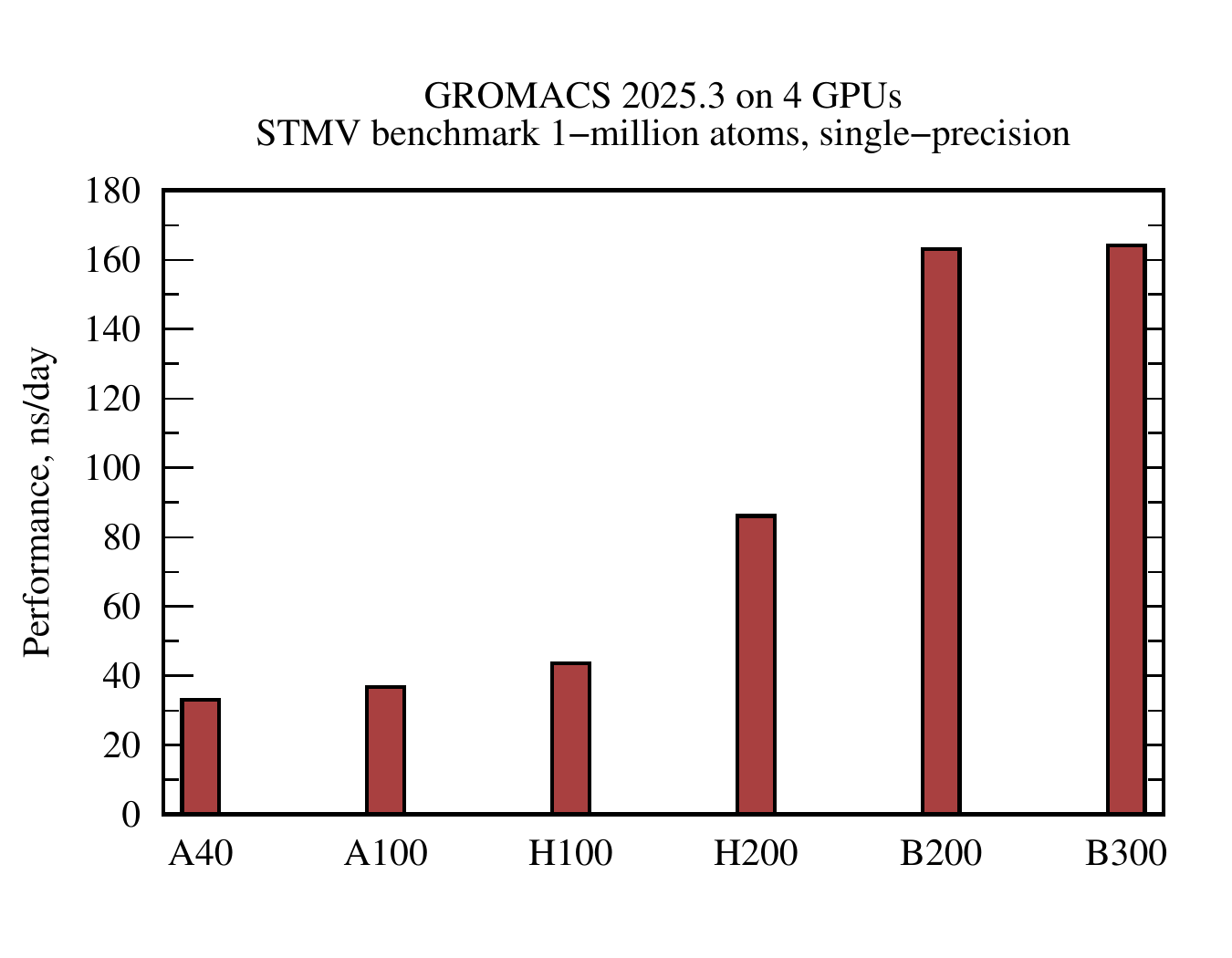}
  \caption{GROMACS 2025.3 on the one-million-atom STMV benchmark, 4 GPUs, single precision throughout. 
  B200 and B300 are within 1\% of each other, 1.9$\times$ the H200 and 4.4$\times$ the A100: single-precision MD is the one class where the two Blackwell parts are interchangeable.}
  \label{fig:gmx}
\end{figure*}

\begin{figure*}[htbp]
  \centering
  \includegraphics[width=0.66\textwidth]{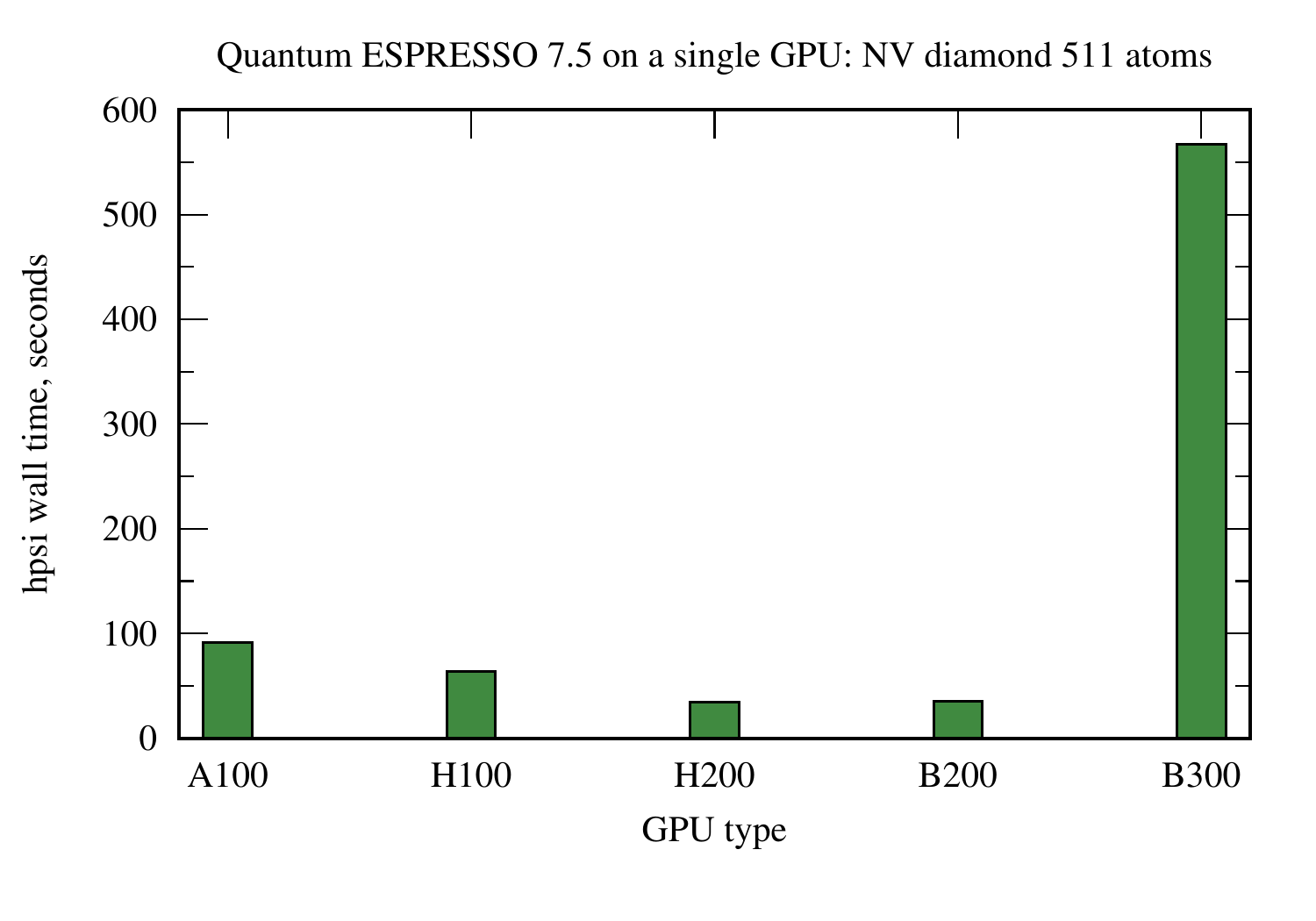}
  \caption{Quantum ESPRESSO 7.5 $G_0W_0$: wall time of the dominant $h\_psi$ kernel on a 511-atom NV-centre diamond supercell, single GPU. \textbf{Lower is better} --- the tall B300
  bar is the 16$\times$ double-precision regression, not a win.}
  \label{fig:qe}
\end{figure*}

These tables and figures give the full HPC detail behind Section~\ref{sec:hpc}:
single-precision molecular dynamics and ML-driven codes (Table~\ref{tab:hpcapps}), NAMD multi-GPU scaling (Table~\ref{tab:namd}), and three application-level figures:
Fig.~\ref{fig:lammps} and Fig.~\ref{fig:qe} carry the double-precision result, while Fig.~\ref{fig:gmx} shows the single-precision case in which the two parts are indistinguishable.

\begin{table}[htbp]
\caption{HPC application performance.}
\label{tab:hpcapps}
\centering
\small
\setlength{\tabcolsep}{4pt}
\begin{tabular}{llrr}
\toprule
Application & Problem & A100 & B200 \\
\midrule
NAMD 3.0 & PCDH15 EC9-MAD12, 67K  & 19 & 42 \\
         & Shaker K$^+$, 202K     & 53 & 118 \\
         & Na/K pump, 333K        & 25 & 68 \\
OpenMM 8.5 & DLPC, 25K            & 81 & 190 \\
         & SRC kinase, 48K        & 222 & 401 \\
         & SERCA1, 277K           & 42 & 84 \\
         & Shaker K$^+$, 202K     & 53 & 104 \\
LAMMPS-MACE & Water, 5174 atoms   & 0.379 & 0.789 \\
Gen-COMPAS & Trp-cage diffusion$^\dagger$ & 30 & 42  \\
\bottomrule
\end{tabular}
\\[2pt]
\footnotesize
MD rows in ns/day. $^\dagger$iterations/s. PCDH15 system from~\cite{pchd15a,pchd15b}.
\end{table}

\begin{table}[htbp]
\caption{NAMD 3.0 GPU-resident mode, 679K atoms, ns/day.}
\label{tab:namd}
\centering
\small
\begin{tabular}{ccrr}
\toprule
GPUs & Optimal threads & B200 & B300 \\
\midrule
1 & 16 & 43.5 & 34.4 \\
2 & 16 & 72.5 & 58.5 \\
4 & 8  & 102.9 & 86.8 \\
\bottomrule
\end{tabular}
\end{table}

\FloatBarrier

\end{document}